\documentstyle[12pt]{article}
 
\begin{document}

\baselineskip=14pt plus 0.2pt minus 0.2pt
\lineskip=14pt plus 0.2pt minus 0.2pt

\newcommand{\be}{\begin{equation}}
\newcommand{\ee}{\end{equation}}
\newcommand{\da}{\dagger}
\newcommand{\dg}[1]{\mbox{${#1}^{\dagger}$}}
\newcommand{\hlf}{\mbox{$1\over2$}}
\newcommand{\lfrac}[2]{\mbox{${#1}\over{#2}$}}

\begin{flushright}
quant-ph/9608008 \\
LA-UR-96-2755 \\
\end{flushright} 

\begin{center}
\large{\bf  
DISPLACEMENT-OPERATOR SQUEEZED STATES.  \\
I.  TIME-DEPENDENT SYSTEMS \\
HAVING ISOMORPHIC SYMMETRY ALGEBRAS \\}
 
\vspace{0.25in}

\large
\bigskip

Michael Martin Nieto\footnote{\noindent  Email:  
mmn@pion.lanl.gov}\\
{\it Theoretical Division, Los Alamos National Laboratory\\
University of California\\
Los Alamos, New Mexico 87545, U.S.A. \\}
 
\vspace{0.25in}

 D. Rodney Truax\footnote{Email:  truax@acs.ucalgary.ca}\\
{\it Department of Chemistry\\
 University of Calgary\\
Calgary, Alberta T2N 1N4, Canada\\}
 
\normalsize

\vspace{0.3in}

{ABSTRACT}

\end{center}
\begin{quotation}

In this paper  
we use the Lie algebra of space-time symmetries to construct 
states which are solutions to the time-dependent Schr\"odinger 
equation for systems with potentials 
$V(x,\tau)=g^{(2)}(\tau)x^2+g^{(1)}(\tau)x+g^{(0)}(\tau)$.   
We describe a set of number-operator 
eigenstates states, $\{\Psi_n(x,\tau)\}$, 
that form a complete set of states but  which, 
however, are usually not  energy 
eigenstates.  From the extremal state, $\Psi_0$, and a 
displacement squeeze operator derived using the Lie 
symmetries, we construct squeezed states and compute expectation 
values for position and momentum as a function of time, $\tau$.  
We prove a general expression for the uncertainty relation for 
position and momentum in terms of the squeezing parameters.  
Specific examples, all corresponding to  choices of 
$V(x,\tau)$ and having isomorphic Lie algebras, will be dealt 
with in the following paper (II).
\vspace{0.25in}


\end{quotation}

\newpage

\baselineskip=.33in

\section{ Introduction}

Recently \cite{I}, we have described the  
unsolved problem of how to define,
for all systems,  generalized squeezed states 
by the displacement-operator method.  As a means to further 
elucidate this problem, we here undertake a study of
systems where there is a Bogoliubov transformation, allowing 
displacement-operator squeezed states to be  
defined.  These states can then be related to the ladder-operator 
squeezed states by  this Bogoliubov transformation.  

Specifically, we will discuss time-dependent systems which
have isomorphic symmetry algebras.  
The isomorphism in the space-time symmetry algebras  
guarantees the existence of transformations which transform the 
time-dependent Schr\"odinger equation for all of these problems 
into a `time-independent' Schr\"odinger equation 
for a one-dimensional harmonic oscillator.  The presence of such  
a transformation means that the Bogoliubov transformation,  
discussed in Reference \cite{I}, exists and the displacement-operator 
squeezed states occur.  

In  the following paper (II) 
we explicitly construct squeezed states for special cases: the 
(well-known) 
harmonic oscillator, the free particle, the linear potential, 
the harmonic oscillator with a uniform driving force, and the 
repulsive oscillator.

In nonrelativistic quantum mechanics, time-dependent systems  
in one spatial dimension can be described by solutions to the 
time-dependent Schr\"odinger equation
\begin{equation}
{\cal S}_1\Psi(x,\tau) = 0,\label{e:I1}
\end{equation}
where the Schr\"odinger operator, ${\cal S}_1$, is 
\begin{equation}
{\cal S}_1 = \partial_{xx} + 2i\partial_{\tau} -2 V(x,\tau).
\label{e:I2}
\end{equation}
The interaction, $V(x,\tau)$, that we will consider here has the  
form
\begin{equation}
V(x,\tau) = g^{(2)}(\tau)x^2 + g^{(1)}(\tau)x + g^{(0)}(\tau),
\label{e:I3}
\end{equation}
where the coefficients, $g^{(j)}(\tau)$, $j=1,2,3$, are differentiable 
and piecewise continuous, but otherwise arbitrary.  We denote the 
solution space of (\ref{e:I1}) by ${\cal F}_{S_1}$.

There are several common problems subsumed by the potential $V(x,\tau)$ 
in Eq. (\ref{e:I3}). We will discuss these individual cases in paper II.  
However, as we will be able to  see in Section 2, all of these  
problems have isomorphic space-time symmetry algebras 
{\cite{drt1,drt2,gt}}.  We will exploit this fact to algebraically
calculate, in Section 3, states of the number operator for 
all such isomorphic systems.

These solution spaces are analogues of the number-operator states of 
the harmonic oscillator {\cite{drt2}} and, in the case of the harmonic 
oscillator, they are indeed the usual number-operator states.  
In addition, for the harmonic oscillator they also correspond to 
the energy eigenstates.  However, in general, for other potentials 
this will not be the case.  Nevertheless, these solution spaces can 
be utilized in the calculation of properties of both coherent states 
{\cite{gt}} and squeezed states, for the general time-dependent 
potential (\ref{e:I3}) and also for the specific cases we will come 
to in paper II.

Coherent {\cite{ks,mmnetal,smsg}} and squeezed states 
\cite{yhhkz}-\cite{mmn2} have received considerable attention in 
the literature in a number of contexts.  In Section 4, we examine 
definitions of squeezed states in the light of the results of the 
Lie symmetry analysis of Section 2.  In Sections 5, we calculate   
expectation values for position and momentum.  We go on, in the 
next section, to obtain the uncertainties in position and momentum, 
and the uncertainty relation when $V(x,\tau)$ is given by Eq. 
(\ref{e:I3}).  


\section{Symmetry}

The generators of space-time symmetries have the general form 
{\cite{drt1,drt2}},
\begin{equation}
{\cal L} = A(x,\tau)\partial_{\tau} + B(x,\tau)\partial_x +  
C(x,\tau).\label{e:II1a}
\end{equation}
For ${\cal L}$ to be a symmetry of Eq. (\ref{e:I1}), then 
${\cal L}\Psi(x,\tau)$ must be a solution of Eq. (\ref{e:I1})  
if $\Psi(x,\tau)$ is a solution.  For this 
to be true, ${\cal L}$ must satisfy the equation {\cite{wm1}}
\begin{equation}
[{\cal S}_1,{\cal L}] =\lambda(x,\tau){\cal S}_1,\label{e:II1b}
\end{equation}
where $\lambda$ is an as yet undetermined function of $x$ and  
$\tau$.  The set 
of all such ${\cal L}$ form a Lie algebra, and the space-time 
Lie symmetry group is obtained accordingly {\cite{wm2}}.

The Lie group of space-time symmetries and its corresponding  
Lie algebra have been identified {\cite{drt1,drt2}} for systems 
with the interaction (\ref{e:I3}). The maximal, complex 
kinematical algebra is $su(1,1)\diamond w^c_1$.  The generators 
of the space-time symmetries have the general form
\begin{eqnarray}
{\cal J}_- & = & \xi\partial_x - ix\dot{\xi} + i{\cal C},\label{e:II19}
\\*[2mm]
{\cal J}_+ & = & \bar{\xi}\partial_x - ix\dot{\bar{\xi}} + i\bar{\cal C}, 
\label{e:II20}\\*[2mm]
I~   & = & 1,\label{e:II21}\\*[2mm]
{\cal M}_- & = & i[\phi_1\partial_{\tau} + (\lfrac{1}{2}\dot{\phi}_1x+
{\cal E}_1)\partial_x-\lfrac{i}{4}\ddot{\phi}_1x^2
-ix\dot{\cal E}_1+\lfrac{1}{4}\dot{\phi}_1+i{\cal D}_1 
+ ig_0\phi_1],\label{e:II22}\\*[2mm]
{\cal M}_+ & = & i[\phi_2\partial_{\tau} + (\lfrac{1}{2}\dot{\phi}_2x+
{\cal E}_2)\partial_x-\lfrac{i}{4}\ddot{\phi}_2x^2
-ix\dot{\cal E}_2+\lfrac{1}{4}\dot{\phi}_2+i{\cal D}_2 + 
ig_0\phi_2],\label{e:II23}\\*[2mm]
{\cal M}_3 & = & i[\phi_3\partial_{\tau} + (\lfrac{1}{2}\dot{\phi}_3x+
{\cal E}_3)\partial_x-\lfrac{i}{4}\ddot{\phi}_3x^2
-ix\dot{\cal E}_3+\lfrac{1}{4}\dot{\phi}_3+i{\cal D}_3 
+ ig_0\phi_3].\label{e:II24}
\end{eqnarray}
The function $\xi$ of $\tau$ and its complex conjugate $\bar{\xi}$ 
are constructed from two real solutions, $\chi_1$ and $\chi_2$, of 
the differential equation
\begin{equation}
\ddot{a}+2g^{(2)}(\tau)a =0.\label{e:II5}
\end{equation}
We choose the Wronskian, $W(\chi_1,\chi_2)=\chi_1\dot{\chi}_2
-\dot{\chi}_1\chi_2 =1$.  The complex solutions of Eq. 
(\ref{e:II5}) are then,  
\begin{equation}
\xi(\tau) =\lfrac{1}{\sqrt{2}}(\chi_1 + i\chi_2),\label{e:II25}
\end{equation}
and its complex conjugate, $\bar{\xi}$.  Their Wronskian is 
\begin{equation}
W(\xi,\bar{\xi}) = \xi\dot{\bar{\xi}} - \dot{\xi}\bar{\xi}=-i.
\label{e:II26}
\end{equation}

We now define the remaining auxiliary $\tau$-dependent functions.
To begin, 
\begin{equation}
{\cal C}(\tau) = \int^{\tau}d\rho\,\xi(\rho)g^{(1)}(\rho)= 
c(\tau)+{\cal C}^o, \label{e:II27}
\end{equation}
where ${\cal C}^o$ is a complex integration constant and the 
function $c(\tau)$ is defined as
\begin{equation}
c(\tau)=\int_{\tau_o}^{\tau}d\rho\,\xi(\rho)g^{(1)}(\rho).
\label{e:II27a}
\end{equation}
We shall take $\tau_o=0$ from this point onward and in paper II.
In addition, we have
\begin{eqnarray}
 & \phi_1(\tau) = \xi^2,~~\phi_2(\tau) = \bar{\xi}^2,~~
\phi_3(\tau) = 2\xi\bar{\xi}, & \label{e:II28}\\*[1.5mm]
 & {\cal E}_1(\tau) = -\xi\left(i{\cal Q}_{1,2}+{\cal C}\right),~~
{\cal E}_2(\tau) = \bar{\xi}\left(i{\cal Q}_{2,1}
-\bar{\cal C}\right), & \nonumber\\*[1.5mm]
 & {\cal E}_3(\tau) = \xi\left(i{\cal Q}_{2,1}-\bar{\cal C}\right) 
-  \bar{\xi}\left(i{\cal Q}_{1,2}+{\cal C}\right), & 
\label{e:II29}\\*[1.5mm]
 & {\cal D}_1(\tau) = -\lfrac{1}{2}\left(i{\cal Q}_{1,2}+
{\cal C}\right)^2,~~
{\cal D}_2(\tau) = -\lfrac{1}{2}\left(i{\cal Q}_{2,1}-
\bar{\cal C}\right)^2, & \nonumber\\*[1.5mm]
 & {\cal D}_3(\tau) =  \left(i{\cal Q}_{1,2}+{\cal
C}\right)\left(i{\cal Q}_{2,1}-\bar{\cal C}\right), & \label{e:II30} 
\end{eqnarray}
where 
\begin{eqnarray}
{\cal Q}_{1,2} & = & \lfrac{1}{2}\xi^o\phi_3^og^{(1)}(0)-q_{1,2}
+i{\cal C}^o, \nonumber\\*[1.5mm]
{\cal Q}_{2,1} & = & \lfrac{1}{2}\bar{\xi}^o\phi_3^og^{(1)}(0) 
- q_{2,1} +i\bar{\cal C}^o,\label{e:II30a}
\end{eqnarray}
and $q_{1,2}$ and $q_{2,1}$ are complex integration constants 
{\cite{kt1}} such that $\bar{q}_{1,2}=q_{2,1}$ and 
$\bar{\cal Q}_{1,2}={\cal Q}_{2,1}$.  

Two integration constants 
appear in each of the equations (\ref{e:II30a}).  Without loss of 
generality, we can choose $q_{1,2}$ such that ${\cal Q}_{1,2}=0$, 
and similarly, we select $q_{2,1}$ such that ${\cal Q}_{2,1}=0$.  
The choices for the remaining integration constants, ${\cal C}^o$ 
and $\bar{\cal C}^o$, will be dictated by the physics of each 
individual system.  Therefore, using Eqs. (\ref{e:II28}) through 
(\ref{e:II30}), we shall drop all references to ${\cal Q}_{1,2}$ 
and ${\cal Q}_{2,1}$.  In addition, we define
\begin{eqnarray}
 & \xi^o=\xi(0),~~\bar{\xi}^o=\bar{\xi}(0),
~~\phi_3^o=\phi_3(0), & \nonumber\\*[1mm]
 & \dot{\xi}^o= \dot{\xi}(0),~~\dot{\bar{\xi}}^o=
\dot{\bar{\xi}}(0),~~
\dot{\phi}_3^o=\dot{\phi}_3(0). & \label{e:II30b}
\end{eqnarray}
Note that $\phi_3^o$ is a real number since $\phi_3$ is a real 
function of $\tau$.

In the general case, 
calculations of expectation values are much simpler 
in terms of the complex algebra and the complex functions in 
Eqs. (\ref{e:II25}) and (\ref{e:II27}) through (\ref{e:II30}).  
However, when working with actual examples, the real functions 
$\chi_1$, $\chi_2$, and the real counterparts of Eqs. 
(\ref{e:II25}) and (\ref{e:II27}) are more advantageous.  
In this paper, we shall use the complex functions.  But in paper 
II,  where we work with specific cases, we will 
transform all equations to expressions in terms of real 
functions.

The operators in Eqs. (\ref{e:II19}) through (\ref{e:II24}) 
satisfy the following (nonzero) commutation relations:
\begin{eqnarray}
 & [{\cal J}_-,{\cal J}_+] = I, & \label{e:II31x}\\*[1.5mm]
 & [{\cal M}_+,{\cal M}_-] = -{\cal M}_3,
~~~~~[{\cal M}_3,{\cal M}_{\pm}] = \pm 2 {\cal M}_{\pm}, & 
\label{e:II32x}\\*[1.5mm]
 & [{\cal M}_3,{\cal J}_{-}] =  -{\cal J}_{-},
~~~~~[{\cal M}_3,{\cal J}_+] = +{\cal J}_+, & 
\label{e:II33ax}\\*[1.5mm]
 & [{\cal M}_{-},{\cal J}_{+}] = -{\cal J}_-,
~~~~~[{\cal M}_+,{\cal J}_{-}]=+{\cal J}_+. & \label{e:II33bx}
\end{eqnarray}
A number of formulae relating the $\tau$-dependent functions 
in Eqs. (\ref{e:II27}) through (\ref{e:II30}) are proven in the 
Appendix.  They are useful in establishing the commutation 
relations (\ref{e:II32x}) to (\ref{e:II33bx}) as well as  
Eq. (\ref{e:II36}) below.

With these commutation relations, we see that the generators 
${\cal J}_{\pm}$ and $I$ form a complexified Heisenberg-Weyl 
algebra, $w_1^c$, and the  operators ${\cal M}_3$ and 
${\cal M}_{\pm}$ close under $su(1,1)$.  Therefore, we have 
the Schr\"odinger algebra in one spatial dimension:
$$({\cal SA})_1^c = su(1,1)\diamond w_1^c.$$

In the following section, we will restrict our analysis to a 
Lie subalgebra of $({\cal SA})_1^c$ consisting of the operators 
${\cal M}_3$, ${\cal J}_{\pm}$, and $I$.  From the 
commutation relations in Eqs. (\ref{e:II31x}) and 
(\ref{e:II33ax}), we recognize that these operators form a 
one-dimensional oscillator algebra, $os(1)$.
It should be noted that the operator ${\cal J}_+$ is the Hermitian 
conjugate of ${\cal J}_-$. Also, $I$ is clearly Hermitian.  
Lastly, the following identity can be demonstrated:
\begin{equation}
{\cal M}_3 = \lfrac{1}{2}\phi_3{\cal S}_1 +{\cal J}_+{\cal J}_-
+\lfrac{1}{2},\label{e:II36}
\end{equation}
This will prove useful in calculating the Casimir operator 
for $os(1)$.


\section{Eigenstates of the Number Operator}

Now we select the operators $\{{\cal M}_3, {\cal J}_{\pm}, I\}$ which 
satisfy the commutation relations
\begin{equation}
[{\cal M}_3,{\cal J}_{\pm}] = \pm {\cal J}_{\pm},
~~[{\cal J}_-,{\cal J}_+] = I.\label{e:III1}
\end{equation}
As mentioned above, we refer to this subalgebra of $({\cal SA})_1^c$ as 
the oscillator subalgebra and denote it by $os(1)$.  It has the 
Casimir operator
\begin{equation}
{\bf C} = {\cal J}_+{\cal J}_- - {\cal M}_3 = 
-\lfrac{1}{2}\phi_3{\cal S}_1 - \lfrac{1}{2}, \label{e:III3}
\end{equation}
which commutes with all the generators in $os(1)$.  The second 
equality in Eq. (\ref{e:III3}) follows from Eq. (\ref{e:II36}).

The fact that all the operators in $({\cal SA})^c_1$ are constants of the 
motion on ${\cal F}_{S_1}$ follows from Eq. (\ref{e:II1b}) 
{\cite{drt2}}.  We select two commuting constants of the motion, 
${\bf C}$ and ${\cal M}_3$, and obtain a set of common eigenvectors.  
Also, we require that these eigenvectors satisfy the  
time-dependent Schr\"odinger equation (\ref{e:I1}) with potential 
(\ref{e:I3}).  The ${\cal J}_{\pm}$ act as ladder operators on the  
eigenvalues of ${\cal M}_3$.  There are three classes of irreducible 
representations of $os(1)$ {\cite{wm3}}.  We are only interested 
in the representation in which the spectrum of ${\cal M}_3$, 
${\rm Sp}({\cal M}_3)$, is bounded below.  Therefore, we have the 
following {\cite{drt2}}:
\begin{eqnarray}
 & {\cal M}_3|m\rangle = (m+\lfrac{1}{2})|m\rangle,~~
{\bf C}|m\rangle = -\lfrac{1}{2}|m\rangle, & 
\label{e:III4}\\*[1.5mm]
 & {\cal J}_+|m\rangle = \sqrt{m+1}|{m+1}\rangle,~~
{\cal J}_-|m\rangle = \sqrt{m}|{m-1}\rangle. & \label{e:III5}
\end{eqnarray}
The condition that the spectrum of ${\cal M}_3$, 
${\rm Sp}({\cal M}_3)$, be bounded below is that 
\begin{equation}
{\cal J}_-|0\rangle =0,\label{e:III5a}
\end{equation}
which defines the extremal state for this representation space.

The states $\Psi_m$ are called number-operator states because they 
are  eigenfunctions of the number operator ${\cal J}_+{\cal J}_-$, 
where
\begin{equation}
{\cal J}_+{\cal J}_-|m\rangle = ({\bf C}+{\cal M}_3)|m\rangle = 
m|m\rangle.\label{e:III6}
\end{equation}
(Note that we go back and forth between  $\Psi_{m}$ and the Dirac-Fock 
notation $|m\rangle$). 
It is important to keep in mind that the generators of $os(1)$ may  
involve an explicit time dependence.  Furthermore, the eigenstates of  
${\cal M}_3$ are solutions to the time-dependent Schr\"odinger 
equation and are not eigenstates of the Hamiltonian except, as 
we shall see, in the case of the 
harmonic oscillator.  Therefore, the number-operator states are not 
generally energy eigenstates.  However, they do provide a convenient, 
complete basis for our purposes {\cite{gt}}.

\indent From Eqs. (\ref{e:III4}) and (\ref{e:III5a}), 
we can calculate the 
specific form of the wave functions.   From the first equation in 
(\ref{e:III4}), we obtain a first-order partial differential equation 
for $\Psi_m$ which can be integrated by the method of characteristics 
{\cite{drt1,kt1,zt}}.  This method leads to ${\cal R}$-separation 
of variables {\cite{wm1}} and yields 
\begin{equation}
\Psi_m(x,\tau) = \exp{\left\{i{\cal R}(x,\tau)\right\}}\psi_m(\zeta)
\Xi_m(\eta), \label{e:III10}
\end{equation}
where the ${\cal R}$-factor is
\begin{equation}
{\cal R}(x,\tau) = \lfrac{1}{4}{{x^2}\over{\phi_3}}\left(\dot{\phi}_3
-\dot{\phi}_3^o\right) 
+{{x}\over{\phi_3^{1/2}}}\left({{{\cal E}_3}\over{\phi_3^{1/2}}}
-{{{\cal E}^o_3}\over{\left(\phi^o_3\right)^{1/2}}}
+\lfrac{1}{2}{\cal B}_3\phi_3^o\right), \label{e:III13}
\end{equation}
and the ${\cal R}$-separable coordinates are
\begin{equation}
\zeta = {{x}\over{\phi_3^{1/2}}}-{\cal B}_3,~~~\eta=\tau.
\label{e:III15}
\end{equation}
The $\eta$-dependent function, $\Xi_m$, is 
\begin{equation}
\Xi_m(\eta) = \left({{\phi_3^o}\over{\phi_3}}\right)^{\lfrac{1}{4}}
\left({{\xi^o\bar{\xi}(\eta)}\over{\bar{\xi}^o\xi(\eta)}}
\right)^{\lfrac{1}{2}\left(m+\lfrac{1}{2}\right)}
\exp{\left[-i\left(\Lambda_3(\eta)+G^{(0)}(\eta)\right)\right]},
\label{e:III17}
\end{equation}
where ${\cal E}_3^o={\cal E}_3(0)$ from Eq. (\ref{e:II29}) 
is a real constant.  The real number $\phi_3^o$ is given in 
Eq. (\ref{e:II30b}).  Furthermore, ${\cal B}_3(\tau)$ is 
defined by the first equality
\begin{equation}
{\cal B}_3(\tau) = \int_{0}^{\tau}ds\,
{{{\cal E}_3(s)}\over{\phi_3^{3/2}(s)}} = b_3(\tau)-b_3(0),
\label{e:III20}
\end{equation}
where
\begin{equation}
b_3(\tau)={{1}\over{\phi_3^{1/2}}}\left[i\left(\xi\bar{\cal C}
-\bar{\xi}{\cal C}\right)\right], \label{e:III23}
\end{equation}
is a real function of $\tau$.  See the Appendix (Formula I) for a 
proof of the second equality in Eq. (\ref{e:III20}).  In addition, 
we define 
\begin{equation}
G^{(0)}(\tau)=\int_{0}^{\tau}ds\,g^{(0)}(s),\label{e:III25}
\end{equation}
and 
\begin{equation}
\Lambda_3(\tau)=\int_{0}^{\tau}ds\,
\left({{{\cal E}_3^2(s)}\over{\phi_3^2(s)}} 
+{{{\cal D}_3(s)}\over{\phi_3(s)}}\right)-{\cal B}_3
\left({{{\cal E}_3^o}\over{\left(\phi_3^o\right)^{1/2}}}+\lfrac{1}{4}
{\cal B}_3^2\dot{\phi}_3^o\right).\label{e:III27}
\end{equation}

Applying ${\cal J}_-$ to $\Psi_0$ from Eq. (\ref{e:III10}) produces a 
first-order ordinary differential equation in $\zeta$ for $\psi_0$.  
Solving this equation leads to a normalized extremal-state wave 
function of the form 
\begin{eqnarray}
\Psi_0(x,\tau) & = & \left({{1}\over{\pi\phi_3^o}}\right)^{\lfrac{1}{4}}
\exp{\left(-b_3^2(0)/2\right)}\exp{\left\{i{\cal R}\right\}}
\exp{\left[-(1-i\theta_1)\zeta^2+(b_3(0)+i\theta_2)\zeta\right]}
\nonumber\\               
 &   & ~~~~~~~~\times \left({{\phi_3^o}\over{\phi_3}}
\right)^{\lfrac{1}{4}}
\left({{\xi^o\bar{\xi}(\eta)}\over{\bar{\xi}^o\xi(\eta)}}
\right)^{\lfrac{1}{4}}
\exp{\left[-i\left(\Lambda_3(\eta)+G^{(0)}(\eta)
\right)\right]},\label{e:III30}
\end{eqnarray}
where
\begin{equation}
\theta_1=\lfrac{1}{2}\dot{\phi}_3^o,~~~~~~\theta_2 = 
{{{\cal E}_3^o}\over{\left(\phi_3^o\right)^{1/2}}}.\label{e:III33}
\end{equation}
The wave function for the state with quantum number $m$ has the form 
\begin{eqnarray}
\Psi_m(x,\tau) & = & \left({{1}\over{m!}}\right)^{\lfrac{1}{2}}
\left({{1}\over{2}}\right)^{\lfrac{m}{2}}
\left({{\bar{\xi}^o}\over{\xi^o}}\right)^{\lfrac{1}{4}}
\exp{\left\{i{\cal R}(x,\tau)\right\}}\psi_m(\zeta)\nonumber\\
               &   & ~~~~~~~~\times 
\left({{\phi_3^o}\over{\phi_3}}\right)^{\lfrac{1}{4}}
\left({{\bar{\xi}}\over{\xi}}\right)^{\lfrac{1}{2}\left(m+\lfrac{1}{2}\right)}
\exp{\left[-i\left(\Lambda_3(\eta)+G^{(0)}(\eta)\right)\right]},
\label{e:III35}
\end{eqnarray}
where
\begin{equation}
\psi_m(\zeta) = H_m(\zeta-b_3(0))
\left({{1}\over{\pi\phi_3^o}}\right)^{\lfrac{1}{4}}
\exp{\left[-\lfrac{1}{2}(1-i\theta_1)\zeta^2+(b_3(\tau_o)+i\theta_2)
\zeta\right]} \label{e:III37}
\end{equation}
and $H_m(\zeta-b_3(0))$ is Hermite polynomial given by the 
Rodrigues formula
\begin{equation}
H_m(\zeta-b_3(0))=(-)^m\exp{\left[\zeta^2-2b_3(0)\zeta\right]}\,
\partial_{\zeta}^m\,\exp{\left[-\zeta^2+b_3(0)\zeta\right]}. 
\label{e:III40}
\end{equation}


\section{ Coherent and Squeezed States}

In notation modified for the present problem, we review the 
general formalism for displacement-operator states.

\subsection{ Coherent states}   
The displacement-operator coherent states {\cite{klauder,amp1}}, 
$\Psi_{\alpha}$, for the systems described by the Schr\"odinger 
equation (\ref{e:I2}) and (\ref{e:I3}), are defined by
\begin{equation}
\vert\alpha\rangle = D(\alpha)\vert 0\rangle,\label{e:IV1}
\end{equation}
where $\alpha$ is a complex number and the displacement operator 
\begin{equation}
D(\alpha) = \exp{(\alpha {\cal J}_+ - \bar{\alpha} {\cal J}_-)},
\label{e:IV2}
\end{equation}
is unitary.  The state $\Psi_0$ is the extremal state (\ref{e:III30}) 
in the number-operator basis, discussed in the previous section.  
Computationally, a more convenient form for the displacement operator 
is given by the expression
\begin{equation}
D(\alpha) = \exp{(-\lfrac{1}{2}|\alpha|^2)}\exp{(\alpha {\cal J}_+)}
\exp{(-\bar{\alpha} {\cal J}_-)}.\label{e:IV3}
\end{equation}

\subsection{ Squeezed states}
  
The generalized squeezed state $\vert {\alpha,z}\rangle$ 
can be obtained from
\begin{equation}
\vert \alpha,z\rangle = D(\alpha)S(z)\vert 0\rangle,
\label{e:IV4}
\end{equation}
where $z$ is a complex parameter and $S(z)$, the squeeze 
operator, is 
\begin{equation}
S(z) = \exp{(z {\cal K}_+ - \bar{z} {\cal K}_-)}.\label{e:IV5}
\end{equation}
The state $\vert 0\rangle$ is the extremal number-operator state 
(\ref{e:III30}).  The operators ${\cal K}_{\pm}$ and ${\cal K}_3$ 
are
\begin{equation}
{\cal K}_- = \lfrac{1}{2}{\cal J}_-^2,~~{\cal K}_+ = 
\lfrac{1}{2}{\cal J}_+^2,
~~{\cal K}_3 = {\cal J}_+{\cal J}_- + \lfrac{1}{2}.\label{e:IV7} 
\end{equation}
These three operators satisfy an $su(1,1)$ Lie algebra with 
commutation relations
\begin{equation}
[{\cal K}_+,{\cal K}_-] = -{\cal K}_3, ~~~~[{\cal K}_3,{\cal K}_{\pm}] 
= \pm {\cal K}_{\pm}. \label{e:IV8}
\end{equation}
Notice the difference between the definition of $K_0$ in reference 
\cite{I} and the operator, ${\cal K}_3$, defined above.  We have 
\begin{equation}
{\cal K}_3 = 2K_0, ~~~~~{\cal K}_{\pm}=K_{\pm}.\label{e:IV7a} 
\end{equation}
This difference is reflected in the commutation relations above 
but does not  affect the remaining calculations in any way.  The 
operators ${\cal K}_-$, ${\cal K}_+$, and ${\cal K}_3$ have the 
important properties 
\begin{equation}
({\cal K}_-)^{\dagger} = {\cal K}_+,~~({\cal K}_+)^{\dagger} 
= {\cal K}_-,~~({\cal K}_3)^{\dagger} = {\cal K}_3, 
\label{e:IV12}
\end{equation}
that is, the operators ${\cal K}_-$ and ${\cal K}_+$ are Hermitian 
conjugates while ${\cal K}_3$ is Hermitian.  Therefore, the squeeze 
operator $S(z)$ is unitary.

The commutation relations of the ${\cal K}_{\pm}$ and 
${\cal K}_3$ with ${\cal J}_{\pm}$ are 
\begin{eqnarray}
 & [{\cal K}_-,{\cal J}_-]=0,~~~[{\cal K}_+,{\cal J}_-]=-{\cal J}_+,~~~
[{\cal K}_3,{\cal J}_-]=-{\cal J}_-, & \nonumber\\*[1mm]
 & [{\cal K}_-,{\cal J}_+]={\cal J}_-,~~~[{\cal K}_+,{\cal J}_+]=0,~~~
[{\cal K}_3,{\cal J}_+]=+{\cal J}_+.\label{e:xAPB2}
\end{eqnarray}

We can express $S(z)$ more conveniently through the 
Baker-Campbell-Hausdorff {\cite{fns,drt3}} relations as
\begin{equation}
S(z) = \exp{(\gamma_+ {\cal K}_+)}\exp{(\gamma_3 {\cal K}_3)}
\exp{(\gamma_- {\cal K}_-)},
\label{e:IV5a}
\end{equation}
where $\gamma_-$, $\gamma_+$, and $\gamma_3$ are analytic  
functions of $z$ and $\bar{z}$
\begin{eqnarray}
 & \gamma_- = -{{\bar{z}}\over{|z|}}\tanh{|z|},~~\gamma_+ = 
{{z}\over{|z|}}\tanh{|z|}, & \nonumber\\*[1.5mm]
 & \gamma_3 = -\ln{(\cosh{|z|})}. & \label{e:IV5b}
\end{eqnarray}
The analytical mappings $\gamma_{\pm}$ and $\gamma_3$ are referred 
to as canonical coordinates of the second kind.  Most of our 
calculations will be carried out with canonical coordinates of the 
second kind.

A definition of squeezed states that is different than 
Eq. (\ref{e:IV4}) can be given by
\begin{equation}
\vert {z,\alpha}\rangle = S(z)D(\alpha)\vert 0\rangle.\label{e:IV6}
\end{equation}
We 
refer to the squeezed state in Eq. (\ref{e:IV4}) as the 
$(\alpha,z)$-representation and to that in (\ref{e:IV6}) as 
the $(z,\alpha)$-representation.  The order of the parameters 
$z$ and $\alpha$ indicates the order the two 
operators $S(z)$ and $D(\alpha)$ have been applied to the extremal 
state.  

Although explicit knowledge of the squeezed-state wave functions 
is not necessary for computation of expectation values of 
functions of position and momentum, it is often important 
to have some representation for them.  One approach is to write 
them as expansions in terms of eigenstates of the 
number operator.  According to Eq. (\ref{e:III30}), the extremal 
state is a Gaussian function.  Starting with the definition 
(\ref{e:IV4}) and the operators (\ref{e:IV3}) for $D(\alpha)$ and 
(\ref{e:IV5a}) for $S(z)$, we have 
\begin{equation}
|\alpha,z\rangle = e^{-\lfrac{1}{2}|\alpha|^2}e^{\alpha{\cal J}_+}
e^{-\bar{\alpha}{\cal J}_-}e^{\gamma_+{\cal K}_+}
e^{\gamma_3{\cal K}_3}e^{\gamma_-{\cal K}_-}|0\rangle.
\label{e:IV50}   
\end{equation}
Given Eq. (\ref{e:III5a}), the definition (\ref{e:IV7}), and the 
fact that ${\cal K}_3|0\rangle =(1/2)|0\rangle$, we obtain 
\begin{equation}
|\alpha,z\rangle = e^{\lfrac{1}{2}(\gamma_3-|\alpha|^2)}
e^{\alpha{\cal J}_+}e^{-\bar{\alpha}{\cal J}_-}
e^{\gamma_+{\cal K}_+}|0\rangle.\label{e:IV52}   
\end{equation}
Next, using the relationship 
\begin{equation}
e^{-\bar{\alpha}{\cal J}_-}e^{\gamma_+{\cal K}_+} = 
e^{(\gamma_+{\cal K}_+-\gamma_+\bar{\alpha}{\cal J}_+)}
e^{-\bar{\alpha}{\cal J}_-},\label{e:IV54}
\end{equation}
and since $[{\cal K}_+,{\cal J}_+]=0$, we find that 
\begin{equation}
|\alpha,z\rangle = e^{\lfrac{1}{2}(\gamma_3-|\alpha|^2)}
e^{\gamma_+{\cal K}_+}e^{(\alpha-\gamma_+\bar{\alpha}){\cal J}_+}
|0\rangle.\label{e:IV56}   
\end{equation}

Expanding the exponentials about the identity, noting Eq. 
(\ref{e:III5}), and using  
$\vert m\rangle=\sqrt{(1/m!)}{\cal J}_+^m|0\rangle$,
we get double summations in terms of the odd and even 
eigenstates
\newpage
\begin{eqnarray}
|\alpha,z\rangle & = & e^{\lfrac{1}{2}(\gamma_3-|\alpha|^2)}\nonumber\\
                 &   & \times \left\{\sum_{m=0}^{\infty}\left[
\sum_{n=0}^{m}\sqrt{{{(2m)!}\over{(2n)!}}}
{{(\alpha-\gamma_+\bar{\alpha})^{2n}\gamma_+^{m-n}}\over{2^{m-n}(m-n)!}}
\right]|2m\rangle\right.\nonumber\\
                 &   & \left.+\sum_{m=0}^{\infty}\left[
\sum_{n=0}^{m}\sqrt{{{(2m+1)!}\over{(2n+1)!}}}
{{(\alpha-\gamma_+\bar{\alpha})^{2n+1}\gamma_+^{m-n}}\over{2^{m-n}(m-n)!}}
\right]|2m+1\rangle\right\},\label{e:IV58}
\end{eqnarray}
where $\gamma_{\pm}$ and $\gamma_3$ are given by Eq. (\ref{e:IV5b}).  
We can derive an expression for $|z,\alpha\rangle$ in a similar 
manner, obtaining 
\begin{eqnarray}
|z,\alpha\rangle & = & e^{\lfrac{1}{2}(\gamma_3+\alpha^2\gamma_-
-|\alpha|^2)}\nonumber\\
                 &   & \times \left\{\sum_{m=0}^{\infty}\left[
\sum_{n=0}^{m}\sqrt{{{(2m)!}\over{(2n)!}}}
{{\alpha^{2n}e^{2n\gamma_3}\gamma_+^{m-n}}\over{2^{m-n}(m-n)!}}
\right]|2m\rangle\right.\nonumber\\
                 &   & \left.+\sum_{m=0}^{\infty}\left[
\sum_{n=0}^{m}\sqrt{{{(2m+1)!}\over{(2n+1)!}}}
{{\alpha^{2n+1}e^{(2n+1)\gamma_3}\gamma_+^{m-n}}\over{2^{m-n}(m-n)!}}
\right]|2m+1\rangle\right\}.\label{e:IV60}
\end{eqnarray}
We shall compare expectation values for the two representations 
of squeezed states in the next section.


\section{Expectation Values for Squeezed States}

In this section we  calculate the expectation values of 
position and momentum in both the $(\alpha,z)$- and the 
$(z,\alpha)$-representations for potentials of the type 
(\ref{e:I3}), where we now use the definitions 
\be
\alpha = |\alpha|e^{i\delta},~~z = re^{i\theta},
~~r=|z|.\label{e:V3}
\ee
We will derive the phase-space trajectories 
for systems with the general potential (\ref{e:I3}).  

Note that 
\begin{eqnarray}
x & = & \bar{\xi}{\cal J}_- + \xi {\cal J}_+ 
+i\left(\xi\bar{\cal C}-\bar{\xi}{\cal C}\right),  
\label{e:V1}\\*[1.5mm]
p & = & \dot{\bar{\xi}} {\cal J}_- + \dot{\xi} {\cal J}_+ + 
i\left(\dot{\xi}\bar{\cal C}-\dot{\bar{\xi}}{\cal C}\right). 
\label{e:V2}
\end{eqnarray}
The proof of Eqs. (\ref{e:V1}) and (\ref{e:V2}) is easily 
demonstrated.  We need only the Wronskian (\ref{e:II26}) and the 
definitions (\ref{e:II19}) of ${\cal J}_-$ and (\ref{e:II20}) 
of ${\cal J}_+$.  We see that 
\begin{eqnarray}
\bar{\xi} {\cal J}_- + \xi {\cal J}_+ & = & x -  
i\left(\xi\bar{\cal C}-\bar{\xi}{\cal C}\right),
\nonumber\\*[1.5mm]
\dot{\bar{\xi}} {\cal J}_- + \dot{\xi} {\cal J}_+ & = & 
-i\partial_{x} - i\left(\dot{\xi}\bar{\cal C}
-\dot{\bar{\xi}}{\cal C}\right).\label{e:APC1}
\end{eqnarray}
By rearranging Eqs. (\ref{e:APC1}), we obtain Eqs. (\ref{e:V1}) 
and (\ref{e:V2}).

We compute the expectation values in both the $(\alpha,z)$- 
and $(z,\alpha)$-representations.  Let ${\cal O}$ be an operator.  
Then, we have the expectation value $\langle {\cal O}\rangle$ 
of ${\cal O}$ in each of the representations
\begin{eqnarray}
\langle {\cal O}\rangle_{(\alpha,z)} 
& = & \langle \alpha,z|{\cal O}|\alpha,z\rangle,  
\nonumber\\*[1.5mm]
 & = &
\langle 0|S^{-1}(z)D^{-1}(\alpha){\cal O}D(\alpha)S(z)
|0\rangle,
\label{e:V4}\\*[2mm] \langle {\cal O}\rangle_{(z,\alpha)} 
& = & \langle
z,\alpha|{\cal O}|z,\alpha\rangle. \nonumber\\*[1.5mm]
& = & \langle
0|D^{-1}(\alpha)S^{-1}(z){\cal O}S(z)D(\alpha)|0\rangle.
\label{e:V5} \end{eqnarray} 
For position and momentum  
operators in the
$(\alpha,z)$-representation,  we have
\begin{eqnarray}
S^{-1}(z)D^{-1}(\alpha)xD(\alpha)S(z) 
& = & X_-(\tau){\cal J}_- + X_+(\tau){\cal J}_+ +  
X_0(\tau)I,\label{e:V6}\\*[1.5mm]
S^{-1}(z)D^{-1}(\alpha)pD(\alpha)S(z) 
& = & \dot{X}_-(\tau){\cal J}_- + \dot{X}_+(\tau){\cal J}_+  
+\dot{X}_0(\tau)I,\label{e:V7}
\end{eqnarray}
where we define the coefficients
\begin{eqnarray}
X_-(\tau) & = &  
\bar{\xi}(e^{\gamma_{3}}-\gamma_-\gamma_+e^{-\gamma_{3}}) 
- \xi \gamma_- e^{-\gamma_{3}},\nonumber\\*[1.5mm]
          & = & \bar{\xi}\cosh{r} + \xi\frac{\bar{z}}{r}\sinh{r},
\nonumber\\*[1.5mm]
          & = & \bar{\xi}\cosh{r} + \xi e^{-i\theta}\sinh{r},
\label{e:V8}\\*[2mm]
X_+(\tau) & = & \xi e^{-\gamma_{3}}+\bar{\xi}\gamma_+  
e^{-\gamma_{3}},
\nonumber\\*[1.5mm]
          & = & \xi\cosh{r} + \bar{\xi}\frac{z}{r}\sinh{r},
\nonumber\\*[1.5mm]
          & = & \xi\cosh{r} + \bar{\xi} e^{i\theta}\sinh{r},
\label{e:V9}\\*[2mm]
X_0(\tau) & = & \alpha\bar{\xi}+\bar{\alpha}\xi 
+i\left(\xi\bar{\cal C}-\bar{\xi}{\cal C}\right), 
\nonumber\\*[1.5mm]
          & = & |\alpha|[e^{i\delta}\bar{\xi}+ 
e^{-i\delta}\xi] +i\left(\xi\bar{\cal C}-\bar{\xi}{\cal C}\right),
\label{e:V10}
\end{eqnarray}
and we have used Eqs. (\ref{e:IV5b}) and (\ref{e:V3}).
In the $(z,\alpha)$-representation, we find that
\begin{eqnarray}
D^{-1}(\alpha)S^{-1}(z)x(\tau)S(z)D(\alpha) & = & 
X_-(\tau) {\cal J}_- + X_+(\tau) {\cal J}_+ + Y_0(\tau)I,
\label{e:V11}\\*[1.5mm] 
D^{-1}(\alpha)S^{-1}(z)p(\tau)S(z)D(\alpha) & = & 
\dot{X}_-(\tau) {\cal J}_- + \dot{X}_+(\tau) {\cal J}_+ 
+ \dot{Y}_0(\tau)I,\label{e:V12}
\end{eqnarray}
where the coefficient $Y_0(\tau)$ is
\begin{eqnarray}
Y_0(\tau) & = & \alpha X_- + \bar{\alpha} X_+  
+i\left(\xi\bar{\cal C}-\bar{\xi}{\cal C}\right),
\nonumber\\*[1.5mm]
          & = & \left(\alpha\bar{\xi}+\bar{\alpha}\xi\right)
\cosh{r}+{{(\alpha\bar{z}\xi+\bar{\alpha}z\bar{\xi})}\over{r}}
\sinh{r}\nonumber\\
          &   & ~~~~~~+i\left(\xi\bar{\cal C}- \bar{\xi}{\cal C}
\right),\nonumber\\*[1.5mm]
          & = & |\alpha|[(\bar{\xi}e^{i\delta} + 
\xi e^{-i\delta})\cosh{r}+(\bar{\xi}e^{i(\theta-\delta)} + \xi
e^{-i(\theta-\delta)})\sinh{r}]\nonumber\\
          &   & ~~~~~~ + i\left(\xi\bar{\cal C}-\bar{\xi}{\cal C}
\right),\label{e:V13} 
\end{eqnarray}
and we have used Eq. (\ref{e:V3}) in the last identity.

Since we have $\langle 0|J_-|0\rangle = \langle 0|J_+|0\rangle = 0$,  
we find that the expectation value for position in the 
$(\alpha,z)$-representation is 
\begin{equation}
\langle x(\tau)\rangle_{(\alpha,z)} =  X_0,\label{e:V14}
\end{equation}
where $X_0$ is given by Eq. (\ref{e:V10}).   
The expectation value for momentum in this representation is
\begin{eqnarray}
\langle p(\tau)\rangle_{(\alpha,z)} & = &  
\dot{X}_0,\nonumber\\*[1.5mm]
                               & = & \alpha\dot{\bar{\xi}}+
\bar{\alpha}\dot{\xi} + i\left(\dot{\xi}\bar{\cal C}
-\dot{\bar{\xi}}{\cal C}\right),\nonumber\\*[1.5mm]
                    & = & |\alpha|[e^{i\delta}\dot{\bar{\xi}}
+ e^{-i\delta}\dot{\xi}] +i\left(\dot{\xi}\bar{\cal C}
-\dot{\bar{\xi}}{\cal C}\right). \label{e:V15} 
\end{eqnarray}

At time $\tau = 0$, let $x_o$ and $p_o$ be the initial position and 
momentum, respectively.  Then, we have
\begin{eqnarray}
 & \langle x(0)\rangle_{(\alpha,z)} = x_o = 
\alpha\bar{\xi}^o + \bar{\alpha}\xi^o+i\left(\xi^o\bar{\cal C}^o
-\bar{\xi}^o{\cal C}^o\right), &  \nonumber\\*[1.5mm]
 & \langle p(0)\rangle_{(\alpha,z)} = p_o = 
\alpha\dot{\bar{\xi}}^o + \bar{\alpha}\dot{\xi}^o 
+ i\left(\dot{\xi}^o\bar{\cal C}^o-\dot{\bar{\xi}}{\cal C}^o\right). 
& \label{e:V16}
\end{eqnarray}
By making use of the Wronskian at $\tau = \tau_o$, 
\begin{equation}
\alpha = i\left(p_o \xi^o - x_o \dot{\xi}^o\right)+i{\cal C}^o.
\label{e:V17}
\end{equation}
Substituting for $\alpha$ and $\bar{\alpha}$ in (\ref{e:V14}) 
and (\ref{e:V15}), we get the general expressions
\begin{eqnarray}
\langle x(\tau)\rangle_{(\alpha,z)} & = & 
i\{[\bar{\xi}(\tau)\xi^o - \xi(\tau)\bar{\xi}^o]p_o +
[\xi(\tau)\dot{\bar{\xi}}^o -\bar{\xi}(\tau)\dot{\xi}^o]x_o\}
\nonumber\\[1.5mm]
                                    &   & 
~~~~~+i\left({\xi}(\tau)\bar{c}(\tau)-{\bar{\xi}}(\tau)c(\tau)\right),
\label{e:V18}\\[2mm]
\langle p(\tau)\rangle_{(\alpha,z)} & = & 
i\{[\dot{\bar{\xi}}(\tau)\xi^o - \dot{\xi}(\tau)\bar{\xi}^o]p_o  
+[\dot{\xi}(\tau)\dot{\bar{\xi}}^o -
\dot{\bar{\xi}}(\tau)\dot{\xi}^o]\}\nonumber\\*[1.5mm]
                                    &   & 
~~~~~+i\left(\dot{\xi}(\tau)\bar{c}(\tau)
-\dot{\bar{\xi}}(\tau)c(\tau)\right), \label{e:V19}
\end{eqnarray}
where $c(\tau)$ is defined by Eq. (\ref{e:II27a}).

The expectation values in the $(z,\alpha)$-representation are  
calculated in a similar way.  For position, we have
\begin{equation}
\langle x(\tau)\rangle_{(z,\alpha)} = Y_0,\label{e:V20}
\end{equation}
where $Y_0$ is given in Eq. (\ref{e:V13}).  For momentum, 
we obtain
\begin{eqnarray}
\langle p(\tau)\rangle_{(z,\alpha)} & = & 
\dot{Y}_0 \label{e:V21o}\\*[2mm]
                                    & = & 
\left(\alpha\dot{\bar{\xi}}+\bar{\alpha}\dot{\xi}\right)\cosh{r} +
\frac{\alpha\bar{z}\dot{\xi}+\bar{\alpha}z\dot{\bar{\xi}}}{r}\sinh{r}
\nonumber\\ 
                                    &   &
 ~~~~~~~~~~+i\left(\dot{\xi}\bar{\cal C}-\dot{\bar{\xi}}{\cal C}\right),
\nonumber\\*[2mm]
                                    & = & 
|\alpha|\left[\left(e^{i\delta}\dot{\bar{\xi}}+e^{-i\delta}\dot{\xi}\right)
\cosh{r} + \left(e^{i(\delta-\theta)}\dot{\xi}+e^{-i(\delta-\theta)}
\dot{\bar{\xi}}\right)\sinh{r}\right]\nonumber\\
                                    &   &
 ~~~~~~~~~~+i\left(\dot{\xi}\bar{\cal C}-\dot{\bar{\xi}}{\cal C}\right).
\label{e:V21}
\end{eqnarray}

\indent From the initial conditions, we get the relationships
\begin{equation}
|\alpha|[e^{i\delta}\cosh{r} + e^{-i(\delta-\theta)}\sinh{r}] =  
i(p_o\xi^o-x_o\dot{\xi}^o)+i{\cal C}^o,\label{e:V22}
\end{equation}
and its complex conjugate.  When these equations are substituted 
into Eqs. (\ref{e:V20}) and (\ref{e:V21}), we obtain results 
which are identical to Eqs. (\ref{e:V18}) and (\ref{e:V19}), 
respectively, in the $(\alpha,z)$-representation.  Since  
the expectation values of position and momentum are identical 
in both the $(z,\alpha)$- and $(\alpha,z)$-representations,  
when we write the expectation values 
of position and momentum in terms of the initial position and 
momentum, we will now drop the representation labels in Eqs. 
(\ref{e:V18}) and (\ref{e:V19}).


\section{Uncertainty Products for Squeezed States}

Next we want to evaluate Heisenberg uncertainty product,  
$(\Delta x)(\Delta p)$, where 
\begin{equation}
(\Delta x)^2 = \langle x^2(\tau)\rangle -\langle x(\tau)\rangle ^2,~~
(\Delta p)^2 = \langle p^2(\tau)\rangle -\langle p(\tau)\rangle  
^2.\label{e:V0}
\end{equation}
In the $(\alpha,z)$-representation the uncertainty in 
position (\ref{e:V0}) can be calculated using (\ref{e:V1}) and 
(\ref{e:V14}):
\begin{equation}
(\Delta x)_{(\alpha,z)}^2 = X_+X_- + X_0^2 - X_0^2 =  
X_+X_-,\label{e:V26}
\end{equation}
where $X_-$ and $X_+$ are given by Eqs. (\ref{e:V8}) and  
(\ref{e:V9}), 
respectively.
In the $(z,\alpha)$-representation we find the same result, since 
\begin{equation}
(\Delta x)_{(z,\alpha)}^2 = X_+X_- + Y_0^2 - Y_0^2 =  
X_+X_-,\label{e:V27}
\end{equation}
where we have employed Eq. (\ref{e:V20}).  Because Eqs. (\ref{e:V26})  
and 
(\ref{e:V27}) are identical, we simply write 
\begin{eqnarray}
(\Delta x)^2 & = & X_+X_-,\label{e:V28}\\*[1.5mm]
             & = & \xi\bar{\xi}\cosh{2r} + 
\lfrac{1}{2}(\bar{\xi}^{\,2}e^{i\theta}+\xi^2e^{-i\theta})\sinh{2r},
\label{e:V28a}  
\end{eqnarray} 
where we have made use of Eqs. (\ref{e:V8}) and (\ref{e:V9}). 
Similarly, we find that the uncertainty in momentum is 
independent of the representation, and we obtain
\begin{eqnarray}
(\Delta p)^2 & = & \dot{X}_+\dot{X}_-,\label{e:V29}\\*[1.5mm]
             & = & \dot{\xi}\dot{\bar{\xi}}\cosh{2r} + 
\lfrac{1}{2}(\dot{\bar{\xi}}^{\,2}e^{i\theta}+\dot{\xi}^2
e^{-i\theta})\sinh{2r}.\label{e:V29a}  
\end{eqnarray}
Therefore, in either representation, the uncertainty relation in  
position and momentum is 
\begin{equation}
(\Delta x)^2(\Delta p)^2 = X_+X_-\dot{X}_+\dot{X}_-.\label{e:V30}
\end{equation}
Substituting for $X_-$ and $X_+$, we have
\begin{eqnarray}
(\Delta x)^2(\Delta p)^2 & = & 
\xi\bar{\xi}\dot{\xi}\dot{\bar{\xi}}\cosh^2{2r}\nonumber\\*[1.5mm]
                         &   & 
~~+
\lfrac{1}{4}({\bar{\xi}}^{\,2}e^{i\theta}
+\xi^2e^{-i\theta})({\dot{\bar{\xi}}}^{\,2}e^{i\theta}+{\dot{\xi}}^2
e^{-i\theta})\sinh^2{2r}\nonumber\\*[1.5mm]
                         &   & ~~-
\lfrac{1}{2}[\bar{\xi}\dot{\bar{\xi}}(\xi\dot{\bar{\xi}}+
\dot{\xi}\bar{\xi})e^{i\theta}+\xi\dot{\xi}(\xi\dot{\bar{\xi}}+
\dot{\xi}\bar{\xi})e^{-i\theta}]\cosh{2r}\sinh{2r},\label{e:V31} 
\end{eqnarray} 
in terms of the complex functions.  

Finally, replacing $\xi$ and $\bar{\xi}$ by Eq. (\ref{e:II25}), 
we obtain an expression for the uncertainty product in terms of 
the real functions, $\chi_1$ and $\chi_2$.  This result,  
\begin{eqnarray}
(\Delta x)^2(\Delta p)^2 & = & 
\lfrac{1}{4}[1+(\chi_1\dot{\chi}_1+\chi_2\dot{\chi}_2)^2]
+\lfrac{1}{8}\Big \{[1+3(\chi_1\dot{\chi}_1+\chi_2\dot{\chi}_2)^2]
\nonumber\\*[1.5mm]
                         &   &
+[(\chi_1\dot{\chi}_1-\chi_2\dot{\chi}_2)^2
-(\chi_1\dot{\chi}_2+\dot{\chi}_1\chi_2)^2]\cos{2\theta}\nonumber
\\*[1.5mm]
                         &   &
~~+2(\chi_1\dot{\chi}_1-\chi_2\dot{\chi}_2)(\chi_1\dot{\chi}_2+
\dot{\chi}_1\chi_2)\sin{2\theta}\Big  
\}\sinh^2{2r}\nonumber\\*[1.5mm]
                         &   &
~~-\lfrac{1}{4}(\chi_1\dot{\chi}_1+
\chi_2\dot{\chi}_2)[(\chi_1\dot{\chi}_1
-\chi_2\dot{\chi}_2)\cos{\theta}
\nonumber\\*[1.5mm]
                         &   &
~~+(\chi_1\dot{\chi}_2+\dot{\chi}_1\chi_2)\sin{\theta}]\sinh{4r},
\label{e:V32}
\end{eqnarray} 
is more revealing and prepares us for paper II.  (Notice that 
when $z=0$, then expression (\ref{e:V31}) or (\ref{e:V32}) 
reduces to the usual uncertainty product for coherent states 
{\cite{gt}}.)


\section*{Acknowledgements}

MMN acknowledges the support of the United States Department of 
Energy.  DRT acknowledges
a grant from the Natural Sciences and Engineering Research Council 
of Canada.


\section*{Appendix}

In this Appendix, we prove four formulae which interrelate  
time-dependent auxiliary functions.  
We refer the reader to Section II for the definitions of the special  
functions required in the proofs.  The first three formulas are 
helpful for calculating the commutation relations of Eqs. 
(\ref{e:II32x}) to (\ref{e:II36}).  Formula IV derives Eq.
(\ref{e:III20}).

\noindent {\bf Formula I.} 

\begin{equation}
\lfrac{1}{2}\dot{\phi_3}b_3(\tau) + {{{\cal E}_3}\over{\phi_3^{1/2}}}  
= i\phi_3^{1/2}(\dot{\xi}\bar{\cal C}-\dot{\bar{\xi}{\cal  
C}}).\label{e:Apx10}
\end{equation}

\noindent {\it Proof}:  Using the definitions of $\phi_3$ and $b_3$, 
we have 
\begin{equation}
\lfrac{1}{2}\dot{\phi_3}b_3 + {{{\cal E}_3}\over{\phi_3^{1/2}}}  
= \lfrac{1}{2}\dot{\phi_3}b_3 + \phi_3\dot{b}_3.\label{e:Apx12}
\end{equation}
\noindent From Eq. (\ref{e:Apx1}), we see that 
\begin{equation}
\dot{b}_3 = -\lfrac{1}{2}{{\dot{\phi}_3}\over{\phi_3}}b_3  
+ 
{{i}\over{{\phi_3^{1/2}}}}.\label{e:Apx14}
\end{equation}
Multiplying by $\phi_3$ and rearranging, we obtain 
\begin{equation}
\lfrac{1}{2}\dot{\phi_3}b_3 + \phi_3\dot{b}_3 =  
{{i}\over{{\phi_3^{1/2}}}},\label{e:Apx16}
\end{equation}
and we are done.
\vskip .5cm

\noindent {\bf Formula II.}

\begin{equation}
{{\ddot{\phi}_3}\over{\phi_3}}  
-\lfrac{1}{2}{{\dot{\phi}_3^2}\over{\phi_3^2}} = -4g_2 +  
{{2}\over{\phi_3^2}}.\label{e:Apx18}
\end{equation}

\noindent {\it Proof}: Substituting for $\phi_3$, we obtain 
\begin{eqnarray}
{{\ddot{\phi}_3}\over{\phi_3}}  
-\lfrac{1}{2}{{\dot{\phi}_3^2}\over{\phi_3^2}} 
& = & 2{{(\ddot{\xi}\bar{\xi}+2\dot{\xi}\dot{\bar{\xi}})}\over{\phi_3}} -  
2{{(\dot{\xi}\bar{\xi}+\xi\dot{\bar{\xi}})^2}\over{\phi_3^2}},
\label{e:Apx20}\\*[1mm]
& = & {{-8g_2\xi\bar{\xi}}\over{\phi_3}} +  
4{{\dot{\xi}\dot{\bar{\xi}}}\over{\phi_3}}  
-2{{(\dot{\xi}\bar{\xi}+\xi\dot{\bar{\xi}})^2}\over{\phi_3^2}},
\label{e:Apx22}\\*[1mm]
 & = & -4g_2 -  
2{{(\xi\dot{\bar{\xi}}-\dot{\xi}\bar{\xi})^2}\over{\phi_3^2}}.
\label{e:Apx25}
\end{eqnarray}
To obtain Eq. (\ref{e:Apx22}) from Eq.  (\ref{e:Apx20}), 
we used the differential 
equation (\ref{e:II5}) for the  
solutions $\xi$ and $\bar{\xi}$.
Then using the Wronskian (\ref{e:II26}) in Eq. (\ref{e:Apx25}), we get 
(\ref{e:Apx18}).  
\vskip .5cm

\noindent {\bf Formula III.}

\begin{equation}
2{{\dot{\cal E}_3}\over{\phi_3}} - {{\dot{\phi}_3{\cal  
E}_3}\over{\phi_3^2}} = -2g_1 - {{2}\over{\phi_3^{3/2}}}b_3.
\label{e:Apx30}
\end{equation}

\noindent {\it Proof}:  Substituting the definitions for 
${\cal E}_3$ and $\phi_3$, we observe that 
\begin{eqnarray}
2{{\dot{\cal E}_3}\over{\phi_3}} - 
{{\dot{\phi}_3{\cal E}_3}\over{\phi_3^2}} 
& = & -2{{(\dot{\xi}\bar{\cal C}+\dot{\bar{\xi}}{\cal C}
+2g_1\xi\bar{\xi})}\over{\phi_3}} +  
2{{(\dot{\xi}\bar{\xi}+\xi\dot{\bar{\xi}})(\xi\bar{\cal C}+\bar{\xi}
{\cal C})}\over{\phi_3^2}},\nonumber\\*[1mm]
& = & -2g_1  
-{{2}\over{\phi_2^2}}[-(\xi\dot{\bar{\xi}}-\dot{\xi}\bar{\xi})\xi
\bar{\cal C} +(\xi\dot{\bar{\xi}}-\dot{\xi}\bar{\xi})\bar{\xi}
{\cal C}],\nonumber\\*[1mm]
& = & -2g_1 - {{2i}\over{\phi_3^2}}(\xi\bar{\cal C}-\bar{\xi}
{\cal C}).\label{e:Apx32}
\end{eqnarray}
When we combine Eq. (\ref{e:Apx1}) with (\ref{e:Apx32}), we  
obtain the desired result.

\vskip .5cm

\noindent {\bf Formula IV,} Eq. (\ref{e:III20}).
\begin{equation}
{\cal B}_3(\tau) = b_3(\tau)-b_3(0),\label{e:Apx1}
\end{equation}
where 
\begin{equation}
b_3(\tau) = {{i(\xi\bar{\cal C} - 
\bar{\xi}{\cal C})}\over{\phi_3^{1/2}}}.\label{e:Apx1a}
\end{equation}

\noindent {\it Proof}:  Recall that we have chosen 
${\cal Q}_{1,2}={\cal Q}_{2,1}=0$.  From 
the definitions of ${\cal E}_3$ and  ${\cal B}_3$ in Eqs. 
(\ref{e:II29}) and (\ref{e:III20}), respectively, we have
\begin{equation}
{\cal B}_3 = \int_0^{\tau}ds\,{{{\cal E}_3}\over{\phi_3^{3/2}}} = 
-\int_0^{\tau}ds\,{{\xi\bar{\cal C}}\over{\phi_3^{3/2}}}  
-\int_{\tau}ds\,{{\bar{\xi}{\cal C}}\over{\phi_3^{3/2}}}.  
\label{e:Apx2}
\end{equation}
Inserting the Wronskian and the definition of $\phi_3$ yields the 
result
\begin{eqnarray}
{\cal B}_3 & = & -\lfrac{i}{2^{3/2}}\int_0^{\tau}ds\,\bar{\cal C}(s)
{{\xi(s)\left(\xi(s)\dot{\bar{\xi}}(s)-\dot{\xi}(s)\bar{\xi}(s)\right)}
\over{\xi(s)^{3/2}\bar{\xi}^{3/2}}}\nonumber\\
           &   & ~~~~~-\lfrac{i}{2^{3/2}}\int_0^{\tau}ds\,{\cal C}(s)
{{\bar{\xi}(s)\left(\xi(s)\dot{\bar{\xi}}(s)-\dot{\xi}(s)\bar{\xi}(s)\right)}
\over{\xi(s)^{3/2}\bar{\xi}^{3/2}}},\nonumber\\*[1mm]
           & = & \lfrac{i}{2^{1/2}}\int_0^{\tau}ds\,\bar{\cal C}(s)\,
d\left({{\xi^{1/2}(s)}\over{\bar{\xi}^{1/2}(s)}}\right)
-\lfrac{i}{2^{1/2}}\int_0^{\tau}ds\,{\cal C}(s)\,
d\left({{\bar{\xi}^{1/2}(s)}\over{\xi^{1/2}(s)}}\right).\label{e:Apx3}
\end{eqnarray}
Integrating by parts, we have 
\begin{eqnarray}
{\cal B}_3(\tau) & = & \lfrac{i}{2^{1/2}}\bar{\cal C}(s)
\left.\left({{\xi^{1/2}(s)}\over{\bar{\xi}^{1/2}(s)}}\right)
\right|_0^{\tau}-\lfrac{i}{2^{1/2}}{\cal C}(s)
\left.\left({{\bar{\xi}^{1/2}(s)}\over{\xi^{1/2}(s)}}
\right)\right|_0^{\tau},\nonumber\\
                 & = & 
\lfrac{i}{2^{1/2}}\left(\bar{\cal C}(\tau)
{{\xi^{1/2}(\tau)}\over{\bar{\xi}^{1/2}(\tau)}}-\bar{\cal C}^o
{{(\xi^o)^{1/2}}\over{(\bar{\xi}^o)^{1/2}}}\right)\nonumber\\
                 &   &  -\lfrac{i}{2^{1/2}}\left({\cal C}(\tau)
{{\bar{\xi}^{1/2}(\tau)}\over{{\xi}^{1/2}(\tau)}}-{\cal C}^o
{{(\bar{\xi}^o)^{1/2}}\over{({\xi}^o)^{1/2}}}\right).\label{e:Apx6}
\end{eqnarray}
Rearranging this expression, we get 
\begin{equation}
{\cal B}_3(\tau) = {{i}\over{\phi_3^{1/2}}}
\left(\xi(\tau)\bar{\cal C}(\tau)-\bar{\xi}(\tau){\cal C}(\tau)\right)
-{{i}\over{(\phi_3^o)^{1/2}}}
\left(\xi^o\bar{\cal C}^o-\bar{\xi}^o{\cal C}^o\right),\label{e:Apx8}
\end{equation}
which is just Eq. (\ref{e:Apx1}) given the definition of $b_3(\tau)$ 
in Eq. (\ref{e:Apx1a}).

\vspace{2mm}


\end{document}